\documentclass{aastex631}

\usepackage{graphicx,epsfig} 	
\usepackage{float}                
\usepackage{amsmath}		
\usepackage{mathtools}	
\usepackage{hyperref} 	
\usepackage{wasysym}
\usepackage{multirow}
\usepackage{array}
\usepackage{upgreek}

\begin{document}

\title{Visual Orbits of Wolf-Rayet Stars II: The Orbit of the Nitrogen-Rich WR Binary WR\,138 measured with the CHARA Array}

\author[0000-0002-0786-7307]{Amanda Holdsworth}
\affiliation{Embry-Riddle Aeronautical University, 3700 Willow Creek Rd, Prescott, AZ 86301, USA}

\author[0000-0002-2806-9339]{Noel Richardson}
\affiliation{Embry-Riddle Aeronautical University, 3700 Willow Creek Rd, Prescott, AZ 86301, USA}

\author[0000-0001-5415-9189]{Gail H. Schaefer}
\affiliation{The CHARA Array of Georgia State University, Mount Wilson Observatory, Mount Wilson, CA 91203, USA}

\author[0000-0002-1722-6343]{Jan J. Eldridge}
\affiliation{Department of Physics, University of Auckland, Private Bag 92019, Auckland 1142, New Zealand}

\author[0000-0002-7648-9119]{Grant M. Hill}
\affiliation{W.M. Keck Observatory, 65-1120 Mamalahoa Highway, Kamuela, HI 96743, USA}

\author{Becca Spejcher}
\affiliation{Department of Physics and Astronomy, Embry-Riddle Aeronautical University, 3700 Willow Creek Rd., Prescott, AZ 86301, USA}

\author[0000-0002-5449-6131]{Jonathan Mackey}
\affiliation{Dublin Institute for Advanced Studies, DIAS Dunsink
Observatory, Dunsink Lane, Dublin 15, D15 XR2R, Ireland}

\author[0000-0002-4333-9755]{Anthony F. J. Moffat}
\affiliation{D\'epartement de physique, Universit\'e de Montr\'eal, Complexe des Sciences, 1375 Avenue Th\'er\`ese-Lavoie-Roux, Montr\'eal, Queb\'ec, H2V 0B3, Canada}

\author{Felipe Navarete}
\affiliation{SOAR Telescope/NSF's NOIRLab, Avda Juan Cisternas 1500, 1700000, La Serena, Chile}

\author[0000-0002-3380-3307]{John D. Monnier}
\affiliation{Astronomy Department, University of Michigan, Ann Arbor, MI 48109, USA}

\author{Stefan Kraus}
\affiliation{Astrophysics Group, Department of Physics \& Astronomy, University of Exeter, Stocker Road, Exeter, EX4 4QL, UK}
\author{Jean-Baptiste Le Bouquin}
\affiliation{Institut de Planetologie et d'Astrophysique de Grenoble, Grenoble 38058, France}
\author{Narsireddy Anugu}
\affiliation{The CHARA Array of Georgia State University, Mount Wilson Observatory, Mount Wilson, CA 91203, USA}
\author{Sorabh Chhabra}
\affiliation{Astrophysics Group, Department of Physics \& Astronomy, University of Exeter, Stocker Road, Exeter, EX4 4QL, UK}
\author{Isabelle Codron}
\affiliation{Astrophysics Group, Department of Physics \& Astronomy, University of Exeter, Stocker Road, Exeter, EX4 4QL, UK}
\author{Jacob Ennis}
\affiliation{Astronomy Department, University of Michigan, Ann Arbor, MI 48109, USA}
\author{Tyler Gardner}
\affiliation{Astrophysics Group, Department of Physics \& Astronomy, University of Exeter, Stocker Road, Exeter, EX4 4QL, UK}
\author{Mayra Gutierrez}
\affiliation{Astronomy Department, University of Michigan, Ann Arbor, MI 48109, USA}
\author{Noura Ibrahim}
\affiliation{Astronomy Department, University of Michigan, Ann Arbor, MI 48109, USA}\author{Aaron Labdon}
\affiliation{European Southern Observatory, Casilla 19001, Santiago 19, Chile}
\author{Cyprien Lanthermann}
\affiliation{The CHARA Array of Georgia State University, Mount Wilson Observatory, Mount Wilson, CA 91203, USA}
\author{Benjamin R. Setterholm}
\affiliation{Astronomy Department, University of Michigan, Ann Arbor, MI 48109, USA}
\affiliation{Max-Planck-Institut für Astronomie, Königstuhl 17, D-69117 Heidelberg, Germany}

\begin{abstract}
Classical Wolf-Rayet stars are descendants of massive OB-type stars that have lost their hydrogen-rich envelopes, and are in the final stages of stellar evolution, possibly exploding as type Ib/c supernovae. It is understood that the mechanisms driving this mass-loss are either strong stellar winds and or binary interactions, so intense studies of these binaries including their evolution can tell us about the importance of the two pathways in WR formation. WR 138 (HD 193077) has a period of just over 4 years and was previously reported to be resolved through interferometry. We report on new interferometric data combined with spectroscopic radial velocities in order to provide a three-dimensional orbit of the system. The precision on our parameters tend to be about an order of magnitude better than previous spectroscopic techniques. These measurements provide masses of the stars, namely $M_{\rm WR} = 13.93\pm1.49M_{\odot}$ and $M_{\rm O} = 26.28\pm1.71M_{\odot}$. The derived orbital parallax agrees with the parallax from \textit{Gaia}, namely with a distance of 2.13 kpc. We compare the system's orbit to models from BPASS, showing that the system likely may have been formed with little interaction but could have formed through some binary interactions either following or at the start of a red supergiant phase, but with the most likely scenario occurring as the red supergiant phase starts for a $\sim 40M_\odot$ star.
\end{abstract}


\section{Introduction} \label{sec:intro}

Classical Wolf-Rayet (WR) stars are evolved massive stars that are core helium-burning and have lost their hydrogen-rich envelope. These stars were first observed at the Paris Observatory by \citet{1867CRAS...65..292W}. There are two evolutionary pathways to create these stars, through single-star or binary-star evolution. In the single star evolutionary pathway, the star lost its envelope through a strong stellar wind in what is now often called the ``Conti scenario" \citep{1975MSRSL...9..193C}. This scenario has strong stellar winds throughout the main-sequence lifetime followed by potential eruptions during a luminous blue variable stage leading to the observed WR star. This scenario may be dominant in some environments, as is evidenced in the recent study of WR stars in the SMC \citep{2024arXiv240601420S}.

The second scenario involves the WR star being formed through interactions with a companion star. In this scenario, the WR star progenitor evolved to fill its Roche lobe and then was stripped of its outer envelope. Recent multiplicity surveys of massive stars have shown that a vast majority of O stars are formed in systems where Roche lobe overflow or mergers may occur for 75\% of O stars \citep{2012Sci...337..444S, 2013A&A...550A.107S, 2014ApJS..215...15S}. This formation mechanism is likely to dominate for WR stars in orbits with periods shorter than $\sim$1 year. However, the exact binary separation or period where the formation channel has to be either through stellar winds with larger separations or binary interactions with smaller orbits remains somewhat ambiguous. For example, \citet{2021MNRAS.504.5221T} studied the massive binary WR\,140 to measure a precise orbit with long-baseline interferometry and optical spectroscopy and compared the results to models from the Binary Population and Spectral Synthesis model grid to show that the WR star in that system formed primarily through mass-loss in the stellar winds. Despite a long 7.93-yr period and a high eccentricity of 0.8993, there was still a moderate amount of mass lost or transferred through binary interactions to form the current system, where the eccentricity could have been the byproduct of imparted ``kicks" near periastron like the models of \citet{2007ApJ...667.1170S}.

Short-period WR binaries are readily studied with photometric, spectroscopic, and polarimetric techniques. However, the amplitudes of variability for all of these techniques become increasingly smaller with longer-period systems. Long-baseline interferometry offers the capabilities of resolving the individual stars moving about each other in longer-period WR binaries that are within $\sim$2-3 kpc. Recently, \cite{Richardson2021} demonstrated that the technique of interferometry could resolve an orbit smaller than 1 mas in separation with the CHARA Array with the first visual orbit of the nitrogen-rich WR binary WR\,133 (WN5o$+$O9I; the ``o" suffix denotes no measurable hydrogen in the WR spectrum). \cite{Richardson2016} resolved two other WR binaries with the CHARA Array, but those observations represent a single epoch and not full orbits. Further observations of WR\,137 have resolved the orbit and helped describe the geometry of the dust formation in the binary (Richardson et al., submitted).

This paper revisits WR\,138 (HD\,193077; WN5o+O9V) which was resolved by \citet{Richardson2016} with the CHARA Array. 
It is one of eight relatively bright WR stars ($V < 8.5$) located in the constellation Cygnus. Although absorption lines in the spectrum of WR\,138 have been recognized by \citep{Hiltner}, the system’s multiplicity remained a topic of debate until recently. 
\citet{Massey} determined that there was no orbital motion for emission lines with an amplitude larger than 30 km s$^{-1}$ over a period of six months during his studies of WR stars with absorption lines present. This led him to suggest that the broad absorption lines, which have an estimated $v \sin i \approx 500$ km s$^{-1}$, were intrinsic to the WN star itself and not formed in the atmosphere of a companion O star.

\citet{Lam} collected a more extensive set of photographic spectra and then 
performed a period search and adopted a period of 2.3238$\pm$0.0001 days as the period of the WN 
suggested that WR 138 is a triple system consisting of the WN6 star orbited by an unseen companion star, potentially a neutron star, every 2.32 days. Both of these objects are orbiting a fainter, rapidly rotating, late-O type main-sequence star every $\sim$1763 days. Following this analysis, \citet{Annuk} collected additional spectra of the system. They found no evidence of the short period suggested by \citet{Lam} and found that the star was a binary with the OB star in a wide orbit with a period of 1538 d (4.2 yr). These results were confirmed by \citet{Palate}, who studied both optical and X-ray data on the system.

\citet{Richardson2016} resolved WR 138 into a binary system using $H-$band CHARA interferometry, deriving a wide separation of 12.4mas, marking the first time a WN binary was resolved with interferometry. They suggested that the system might have gone through a previous mass-transfer episode, which created the WR star and presented a spectral model of the system using the non-LTE code PoWR, measuring the mass-loss rates and properties of the two stars in the system. Inspired by the stars being resolved with CHARA, \citet{Rauw} examined spectra taken of the system spanning nearly three orbits. 
They also confirmed that there is no signal in the radial velocity time series at frequencies around the 2.3238 day period found by \citet{Lam}.
After analyzing the results provided by \citet{Richardson2016}, \citet{Rauw} identified discrepancies between the CHARA observations and their own spectroscopic radial velocity solution. They suggested that the secondary star resolved through interferometry was not responsible for the orbital motion of the WN6o star with a period of 1559 days, but rather the interferometric companion was a putative third component that does not undergo significant RV variations.

The aim of our study is to better characterize the WR 138 binary system, determining the orbital parameters and masses of both stars through a combination of spectroscopy and new interferometry from the CHARA Array. We present the observations in Section 2, along with our astrometric and spectroscopic measurements in Section 3. Then, in Section 3, we present the 3-dimensional orbit. We discuss our results in Section 4 and conclude this study in Section 5.

\section{Observations}

\subsection{Long-baseline infrared interferometry with the CHARA Array}

Following the tentative detection of the resolved binary by \citet{Richardson2016}, we began a long-term program with the CHARA Array \citep{2005ApJ...628..453T} to measure the orbital motion of the system with long-baseline interferometry. We collected data with the CLIMB beam combiner \citep{2013JAI.....240004T} both with the observations reported by \citet{Richardson2016} taken in 2013 August and on three additional epochs in 2018 June, July, and August. These new CLIMB measurements, as well as the data published by \citet{Richardson2016} suffer from poor $(u,v)$ coverage and limited measurements of the squared visibility ($V^2$) and closure phases ($CP$). 

In 2019, our program began using measurements with the Michigan InfraRed Combiner - eXeter (MIRC-X) beam combiner on the CHARA Array \citep{2020AJ....160..158A}. This instrument utilizes up to all six telescopes of the Array and was an upgrade to the four- and then six-telescope combiner MIRC \citep{2006SPIE.6268E..1PM, 2012SPIE.8445E..0ZC}. MIRC-X was used with the PRISM50 mode, allowing for 8 spectral channels across the $H-$band, with a spectral resolving power of $R\sim50$. Often the spectral channels at the edges of the $H-$band are rejected due to low signal-to-noise, meaning we end up with 6 spectral channels in each data-set. Unlike the CLIMB data, the $(u,v)$ coverage was much improved with a single observation and the resulting measurements of $V^2$ and $CP$ of exceptional quality resulting in measurements of the separation and position angle with precision close to 10 $\upmu$as (see Section 3). 

In August 2021, the CHARA Array commissioned a second six-telescope beam combiner, the Michigan Young Star Imager at CHARA \citep[MYSTIC;][]{2023JATIS...9b5006S}. MYSTIC observes in the $K$-band and operates simultaneously with MIRC-X. We used MYSTIC in PRISM49 mode, providing 11 spectral channels across the $K-$band with a spectral resolving power of $R\sim50$. Similarly to MIRC-X, the channels at the edges of the bandpass are often rejected, leaving us with nine useful wavelength channels across the $K-$band. 

All MIRC-X and MYSTIC data were reduced using the pipeline\footnote{\href{https://gitlab.chara.gsu.edu/lebouquj/mircx\_pipeline}{https://gitlab.chara.gsu.edu/lebouquj/mircx\_pipeline}} (version 1.3.3–1.3.5) developed by Jean-Baptiste Le Bouquin and the MIRC-X team \citep{le_bouquin_2024_12735292}, which splits each 10-minutes data sequence into four 2.5-minute bins. These reductions produce squared visibilities ($V^2$) for each baseline and $CP$s for each closed triangle of telescopes.
The use of 6 telescopes simultaneously allows for measurements of the squared visibility across 15 baselines with a simultaneous measurement of 20 different closure phases. We show the calibrators used and their diameters in Table~\ref{calibrators}.

\begin{table*}
\centering \rotate
\caption{Calibrator stars observed during the MIRC-X, and MYSTIC observations at the CHARA Array. A $\checkmark$ denotes the night this star was used as a calibrator. Calibrators found from the JMMC SearchCal database \citep{2006A&A...456..789B, 2011A&A...535A..53B}. \label{calibrators}}
\begin{tabular}{lcccccc}
\hline \hline
Calibrator star	&	$\theta_{{\rm UD},H}$ (mas)	&	$\theta_{{\rm UD},K}$ (mas)	&	2019 Jul 01	&	2019 Jul 02 	&	2019 Sep 5	&	2021 Aug 02		\\ \hline
HD\,178538	&	0.248715	&	0.249373	&	$\checkmark$	&	$\checkmark$ &	$\checkmark$	&	$\checkmark$				\\
HD\,191703	&	0.218459	&	0.219038	&	$\checkmark$	&	$\checkmark$ &	$\checkmark$	&				\\
HD\,192536	&	0.166190	&	0.166553	&		&		&		&			\\
HD\,201614	&	0.317421	&	0.318844	&	$\checkmark$	&	&	$\checkmark$	&	$\checkmark$	\\
HD\,197176	&	0.241453	&	0.242173	&	$\checkmark$	&	$\checkmark$ &	$\checkmark$	&	$\checkmark$				\\
HD\,192732	&	0.400280	&	0.402075	&		&	 &		&					\\
\hline															
(continued) &	2021 Oct 22	 &	2022 Jul 19	&	2022 Aug 23	&	2023 Jun 03	&	2023 Aug 14 & \\ \hline 
HD\,178538	&	 & $\checkmark$	&		&	$\checkmark$	&	$\checkmark$	& \\
HD\,191703	&	$\checkmark$	& $\checkmark$	&	$\checkmark$	&	$\checkmark$	&	$\checkmark$	&  \\
HD\,192536	&	& $\checkmark$	&		&		&	$\checkmark$ & 	\\
HD\,201614		& $\checkmark$ &		&		&		&	$\checkmark$	& \\
HD\,197176	& &	$\checkmark$	&		&	$\checkmark$	&	$\checkmark$	& \\
HD\,192732	&	&	&	$\checkmark$	&		&	  & \\
\hline \hline
\end{tabular}
\end{table*}

For each MIRC-X/MYSTIC night, we compared the calibrators against each other and found no evidence for binarity after visually inspecting the data, allowing us to know that our calibrations and subsequent binary measurements were of high quality. We applied wavelength correction factors by dividing the wavelengths in the MIRC-X OIFITS files by a factor of 1.0054 $\pm$ 0.0006 and those in the MYSTIC OIFITS files by 1.0067 $\pm$ 0.0007 \cite[][Monnier, priv. comm]{2022AJ....164..184G}. 

\subsection{Spectroscopy}

Many of our spectroscopic measurements were taken from previously published data for the orbit of the system. Our team has also collected spectra from the Dominican Astronomical Observatory 1.8-m telescope (DAO), with a resolving element of 1.33 angstroms over a span of approximately 27 years. Unfortunately, the resolving power of $\approx3800$ near blaze maximum ($\sim5600$\AA), yielded measurements that were not precise enough to better constrain the orbital motion and hence not used here. 
We also used the Keck observatory and the Echellette Spectrograph and Imager \citep[ESI;][]{2002PASP..114..851S}. These data were taken over a range of three years. These four spectra have a resolving power of nearly 10,000, with a typical signal-to-noise ratio (SNR) of 120.
To better constrain the motion of both stars, we also used the data from \citet{DSilva}, which has a resolving power of $R\sim 85,000$ and a typical SNR of 75. We summarize our spectroscopic measurements in Table \ref{spectra}. We did not use the more limited data sets reported by \citet{Massey} or \citet{Lam} due to their coverage not spanning a full orbit of the system, along with the photgraphic plate measurements having larger errors. The measurements from \citet{Annuk} did not include details like resolving power or the full wavelength range, where it is just listed as ``blue" but the emission line kinematics were from \ion{N}{4} 4057 and \ion{He}{2} 4686. 

\begin{table*}[]
    \centering
    \begin{tabular}{l c c c c c c}
    \hline \hline 
        Telescope & Spectrograph & Resolving & $N_{\rm spec}$ & Wavelength  & Date Range & Ref. \\
        & & Power & & Range ($\AA$) & (HJD-2440000)  & \\
        \hline 
        Tartu 1.5 m & Cassegrain & \ldots & 73 & \ldots & 4485.371 - 7029.470 & \citet{Annuk}
        \\
        Mercator 1.2 m & HERMES & $85,000$ & 40 & 4000 - 9000 & 16126.456 - 19024.6 &\citet{DSilva} 
        \\ 
        OHP & Aurélie & $10,000$ & 8 & 4448 - 4886 & 15827.801 - 19853.787 & \citet{Rauw} 
        \\
        TIGRE & HEROS & $20,000$ & 37 & 3760 - 8700 & 17508.938 - 20043.958 & \citet{Rauw}
        \\
        KECK & ESI & $8,829$ & 4 & 5200 - 6000 & 19024.106 - 20150.093 & This paper   
        \\
        \hline \hline 
    \end{tabular}
    \caption{Spectroscopic data sets used in our analysis. The details are given in the text, with some of these being used with the published measurements only.}
    \label{spectra}
\end{table*}

\section{Measurements and determination of the orbit}

\subsection{Astrometric measurements with the CHARA Array}

We follow the procedures developed by \citet{2016AJ....152..213S} to fit the interferometric measurements of WR 138 made with the CHARA Array, as has been done in past orbits of WR stars measured with CHARA \citep{Richardson2016, Richardson2021, Thomas}. The binary positions are fitted using a grid-search code\footnote{The code is available at \href{https://www.chara.gsu.edu/analysis-software/binary-grid-search/}{https://www.chara.gsu.edu/analysis-software/binary-grid-search/}.}. This code uses both $V^2$ and $CP$, which helps to remove a 180$^\circ$ ambiguity from the position angle. { The visibility amplitudes measure the size and shape of the source, while the closure phases are sensitive to asymmetries in the light distribution. Both the $V^2$ and $CP$ are used to measure the binary separation and flux ratio.
 With each fit,} there are two options based on whether we have the brighter or fainter star as the central star. The approach calculates a $\chi^2$ statistic for the data based on a binary model for a large grid of separations in right ascension and declination. At each step in the grid, the IDL mpfit package \citep{2009ASPC..411..251M} is used to optimize the binary position and flux ratio between the two stars. The global minimum across the grid is selected as the best fit solution. We did a thorough search by varying the separations in increments of 0.5 mas across a range of $\pm20$ mas in { the directions of} both $\triangle$RA and $\triangle$DEC. The resulting plots are shown in a figure set in the appendix. The $\chi^2$ maps for the CLIMB data had many local minima and were not consistent with the measurements made with MIRC-X and MYSTIC and thus we did not include them in our analysis.
Furthermore, this can explain the inconsistencies pointed out by \citet{Rauw}. Our measured separations, position angles, error ellipses, and flux ratios are presented in Table \ref{interferometry}.

\begin{table*}
\centering
\caption{Interferometric measurements of the binary with the CHARA Array. 
\label{interferometry}}
\begin{tabular}{l c c c c c c c c c c}
  \hline \hline
UT Date & HJD 	&	Filter	&	Position	&	Separation	&	$\sigma_{\rm major}$	&	$\sigma_{\rm minor}$	&	$\sigma_{\rm PA}$	&	$f_{\rm WR}$	&	$f_{\rm O}$	& Comb.	\\	
 & $-$2,400,000	 &		&	Angle ($^\circ$)	&	(mas)	&	(mas)	&	(mas)	&	($^\circ$)	&		&		\\	\hline

2019 July 1	    &  58665.772	&	$H$	 &	122.161	&	3.8777	&	0.0208	&	0.0134	&	1.29	& 0.65 & 0.35 & 	M	\\
2019 July 1	    &  58665.972	&	$H$	 &	122.405	&	3.902	&	0.0083	&	0.0049	&	139.78	& 0.63 & 0.37 & 	M	\\
2019 July 2     &  58666.807	&	$H$	 &	122.577	&	3.9013	&	0.0102	&	0.0081	&	146.34	& 0.63 & 0.37 & 	M	\\
2019 July 2     &  58666.995	&	$H$	 &	122.666	&	3.9041	&	0.0093	&	0.0051	&	49.33	& 0.61 & 0.39 &  	M	\\
2019 September 5 & 58731.867	&	$H$	 &	124.337	&	3.8757	&	0.0095	&	0.0046	&	70.78	& 0.65 & 0.35 &   	M	\\
2021 August 2    & 59428.887	&	$H$	 &	306.829	&	4.1585	&	0.0079	&	0.0051	&	52.47	& 0.62 & 0.38 & 	M	\\
2022 July 19     & 59779.982	&	$H$	 &	346.235	&	0.7356	&	0.0076	&	0.0042	&	62.21	& 0.64 & 0.36 & 	M	\\
2022 July 19     & 59779.982	&	$K$	 &	347.508	&	0.726	&	0.0088	&	0.0035	&	77.66	& 0.69 & 0.32 & 	Y	\\
2022 August 23   & 59814.812	&	$H$	 &	28.815	&	0.4985	&	0.0090	&	0.0039	&	95.25	& 0.59 & 0.41 & 	M	\\
2022 August 23   & 59814.812	&	$K$	 &	30.42	&	0.5233	&	0.0176	&	0.0109	&	92.08	& 0.65 & 0.35 & 	Y	\\
2023 June 3      & 60098.817	&	$H$	 &	119.549	&	3.4243	&	0.0074	&	0.0032	&	129.21	& 0.64 & 0.36 & 	M	\\
2023 June 3      & 60098.817	&	$K$	 &	119.595	&	3.4272	&	0.0213	&	0.0141	&	136.16	& 0.70 & 0.30 & 	Y	\\
2023 August 14	 & 60170.768	&	$H$	 &	121.824	&	3.8397	&	0.0079	&	0.0053	&	119.48	& 0.66 & 0.34 & 	M	\\
2023 August 14	 & 60170.768	&	$K$	 &	121.731	&	3.832	&	0.0095	&	0.0072	&	112.47	& 0.70 & 0.30 & 	Y	\\ \hline 
\hline
\end{tabular}
\tablecomments{M = MIRC-X, Y = MYSTIC}
\end{table*}

From these measurements, we were able to fit a visual orbit following the procedures\footnote{Available at \href{http://www.chara.gsu.edu/analysis-software/orbfit-lib}{http://www.chara.gsu.edu/analysis-software/orbfit-lib}} of \citet{Schaefer2006, 2016AJ....152..213S}. With the formal errors from the binary position fits, we found an orbit with a period of 1529.3$\pm$1.5 d, $e = 0.191\pm 0.004$, $a = 4.172\pm0.007$ mas, and an inclination of 84.21$\pm$0.05$^\circ$. The reduced $\chi^2$ statistic from this fit had a value of 5.37, so we scaled the uncertainties to have a visual orbit fit have a reduced $\chi^2$ statistic of unity to help account for systematic errors in our data. We report these scaled error ellipses in Table \ref{interferometry}. 

\subsection{Spectroscopic Measurements}

In order to best fit the orbit of WR 138, we also wanted to incorporate the spectroscopic measurements of the stars into our fit. We began by trying a combined fit of the SB2 and visual orbit with the velocities from \citet{Rauw}. We found that the low-amplitude values of the O star caused the orbit fitting routines to produce results that were not fitting the orbit compared with those of either \citet{Rauw} or our visual orbit. Given \citet{Rauw} did not fit an SB2 directly, but rather fit the WR component and then used a linear regression of the velocities of the two stellar components to infer a mass ratio in the system. This is justified here as the noise in the radial velocities of the O star measurements is large enough to prevent a good fit to the O star velocities. Therefore, we began our work by doing a visual and SB1 (Wolf-Rayet component) combined fit with the WR velocities reported by \citet{Rauw}. 

\begin{figure}
    \centering
    \includegraphics[width=3.5in]{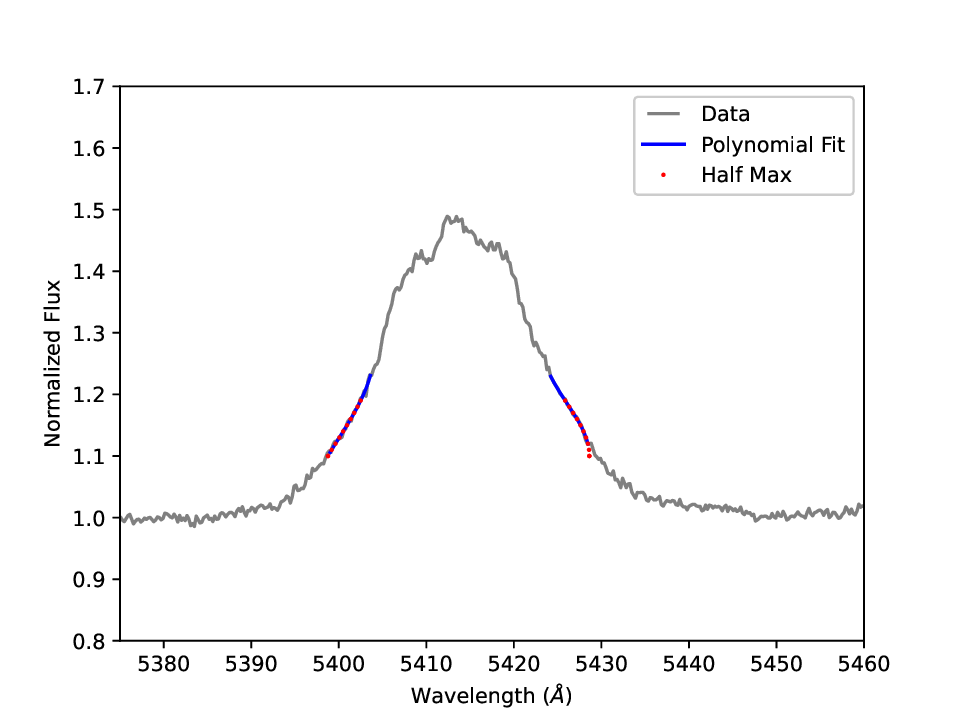}
    \includegraphics[width=3.5in]{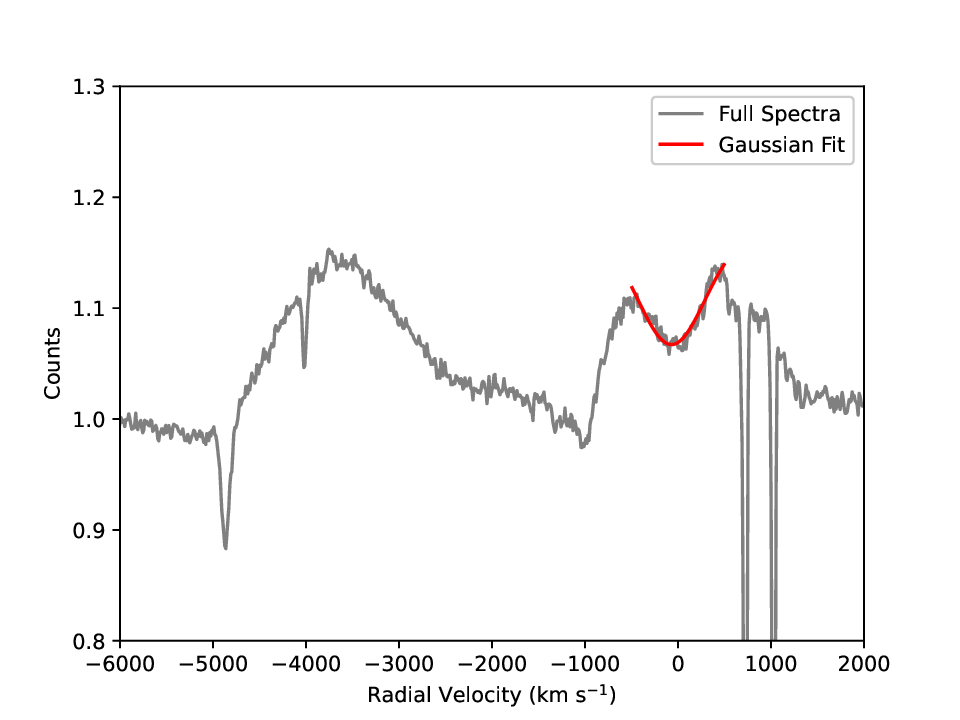}
    \caption{Example Bisector Fit from the ESI data set, HJD: 2459371 of the \ion{He}{2} $\lambda5411$ emission line (left) and the Gaussian fit of the \ion{He}{1} $\lambda5876$ absorption line (right). The grey line represents the spectrum, with the blue line showing where the bisector fit was taken. The red dots along the blue lines show where the measurements were taken in order to average and get the center radial velocity. The red line in the right panel indicates the Gaussian fit to the O star absorption line.}
    \label{fig:BisectorFit}
\end{figure}

Once this orbital fit was successful, we combined in other data sets. In addition to the data from \citet{Rauw}, we measured the spectra from \citet{DSilva} and included the measurements of \citet{Annuk}. We also used the spectra we collected with Keck and the ESI. To measure the WR star's velocity, we used the bisector technique for emission lines that has been used for many WR stars, with methods and code documented recently by \citet{2023MNRAS.519.5882S}. We show the bisector of an example spectrum around \ion{He}{2} $\lambda5411$ in Fig.~\ref{fig:BisectorFit}. The O star's velocities were measured using a Gaussian fit to the \ion{He}{1} $\lambda5876$ line, on the absorption between the radial velocity range of -500 and 500 km s$^-1$ (also shown in Fig.~\ref{fig:BisectorFit}). We used the dispersion of the velocities in the interstellar \ion{Na}{1} D lines to gauge the accuracy and precision of the wavelength calibrations of each spectrum.

\begin{figure}
    \centering
    \includegraphics[width=3.5in]{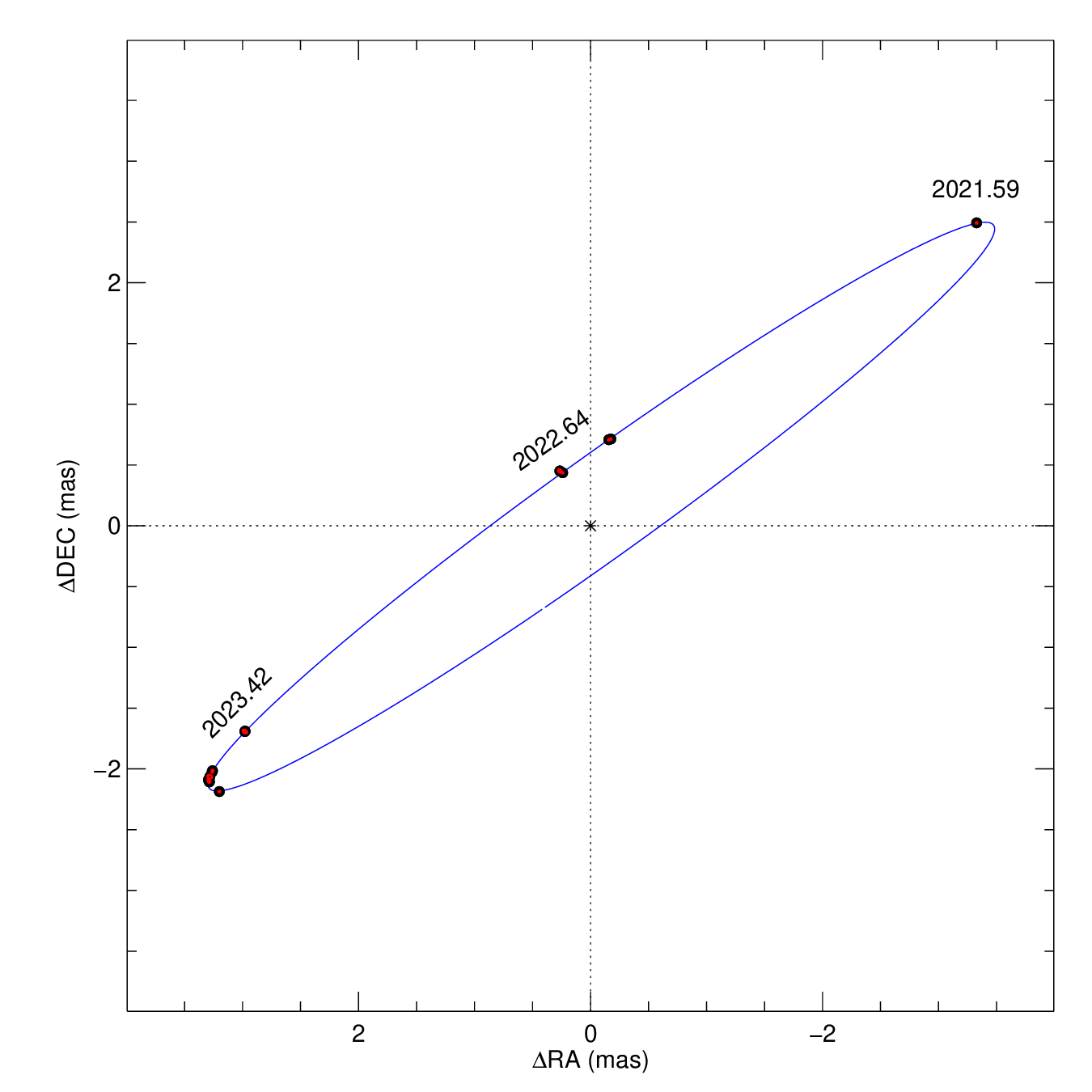}
    \includegraphics[angle=90,width=3.5in]{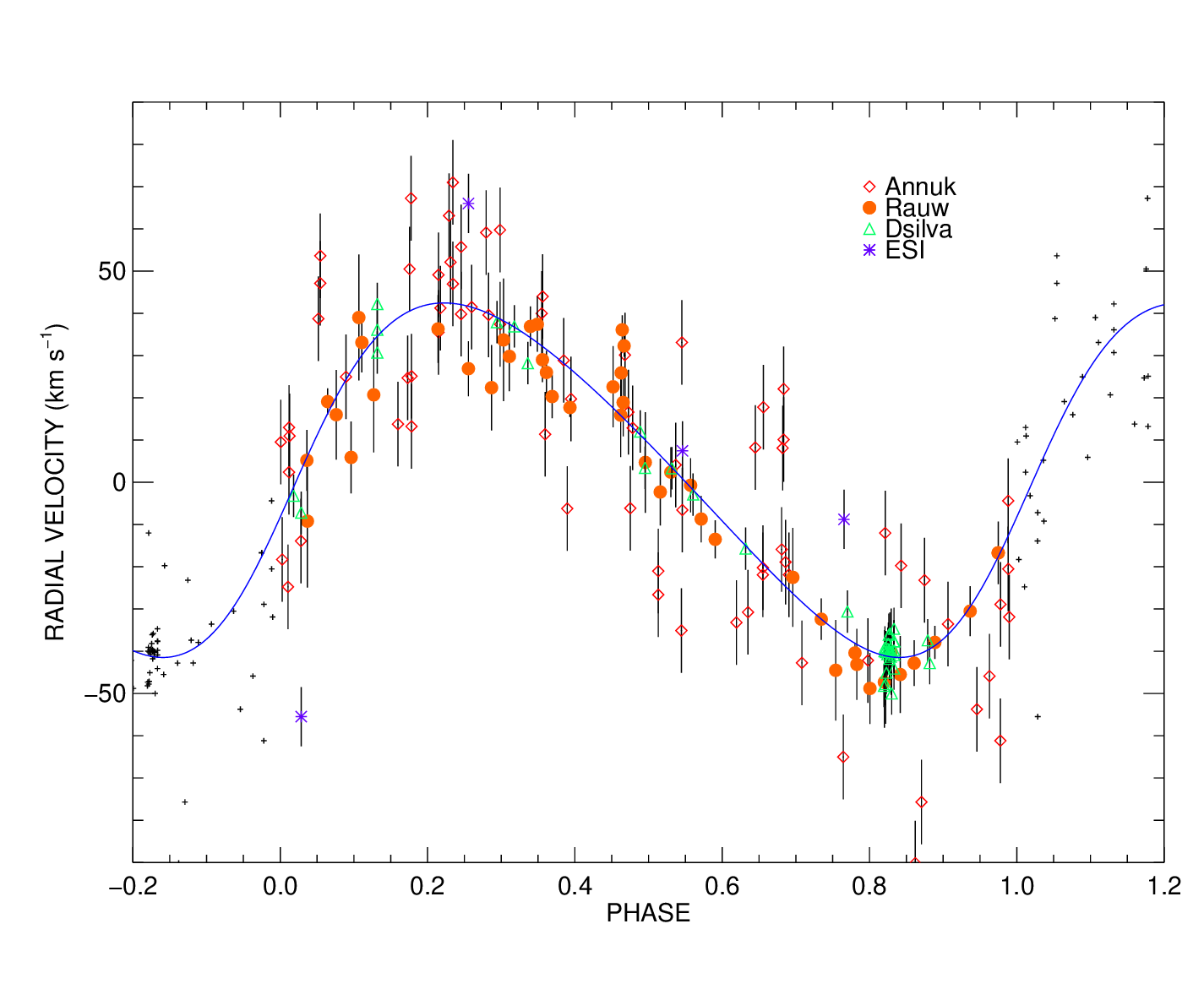}
    \caption{The orbital solution for the CHARA measurements (left), along with the measurements of the WR star's radial velocities (right). The visual orbit model is in blue, with the black points representing the measurements of the O star moving around the WR star. Red dots show the measurement errors of the interferometry.}
    \label{fig:visorbit}
\end{figure}

We found that the combined visual and single-lined spectroscopic fit still had a large scatter in the radial velocity orbit. In order to minimize the scatter, we used the derived orbital parameters to fit each subset of spectra with the same orbital elements varying only the $\gamma$-velocity. We then adjusted each subset of spectra to have the same $\gamma$-velocity of the data from \citet{Rauw}. A third body could explain a change in the $\gamma$-velocity, but each dataset was measured differently and WR stars are difficult to measure radial velocities to high precision. Therefore, we do not report a $\gamma$-velocity in Table \ref{OrbitalElements}. With this larger dataset of spectroscopy, we were able to do a combined fit of the visual and spectroscopic orbit of the WR star. The orbital elements are presented in Table \ref{OrbitalElements} and the fits are shown in Fig.~\ref{fig:visorbit}. We compare the orbital elements of both \citet{Annuk} and \citet{Rauw} in Table \ref{OrbitalElements}, which shows both our higher precision and highlights our ability to measure properties of the orbit such as the inclination and hence stellar masses.

\begin{table}[h]
\centering
\caption{Orbital Elements \label{OrbitalElements}}
\begin{tabular}{l c c c}
 \hline
 \multicolumn{2}{c}{Measured Quantities} \\
 \hline
 Orbital Element & Value & \citet{Annuk} & \citet{Rauw}\\
 \hline
 $P$ (d) & 1527.99 $\pm$ 1.01       & 1538      & 1553$\pm$14\\
 $P$ (yr) & 4.18 $\pm$ 0.003  &    4.21  &  4.25$\pm$0.04 \\
 $T_0$ (MJD) & 52868.62 $\pm$ 4.98  & 45284$\pm$39 & 57343$\pm$66 \\
 $T_0$ (yr) & 2003.629 $\pm$ 0.013  &  1982.86$\pm$0.11     &  2015.88$\pm$0.18 \\
 $e$ & 0.191 $\pm$ 0.0046  &    0.29$\pm$0.05   & 0.15$\pm$0.04 \\
 $a$ (mas) & 4.17 $\pm$ 0.0087  & $\cdots$ & $\cdots$ \\
 $i$ & 84.21 $\pm$ 0.06& $\cdots$ & $\cdots$\\
 $\Omega$ ($^{\circ}$) & 124.32 $\pm$ 0.05 & $\cdots$ & $\cdots$\\
 $\omega_{WR}$ ($^{\circ}$) & 258.00 $\pm$ 0.36 & 271$\pm$12 & 233$\pm$16\\
 $K_1$ (km s$^{-1}$) & 41.95 $\pm$ 0.904 & 30.6$\pm$1.9 & \\
 \hline
 \multicolumn{2}{c}{Derived Quantities} \\
 \hline
 Quantity & Fit \\
 \hline
 $M_{WR}$ ($M_{\odot}$) & 13.93 $\pm$ 1.49& $\cdots$ & $\cdots$\\
 $M_O$ ($M_{\odot}$) & 26.28 $\pm$ 1.71 & $\cdots$ & $\cdots$\\
 $a_1$ (AU) & 5.81 $\pm$ 0.125 & $\cdots$ & $\cdots$\\
 $a_2$ (AU) & 3.08 $\pm$ 0.188 & $\cdots$ & $\cdots$\\
 $d$ (pc) & 2131.97 $\pm$ 54.38 & $\cdots$ & $\cdots$\\
 Parallax (mas) & 0.469 $\pm$ 0.012 & $\cdots$ & $\cdots$\\
 \hline
\end{tabular}
\end{table}

To obtain a full three-dimensional orbit, we needed the semi-amplitude of the O star's orbit as well. We compared the O and WR star velocities and performed a linear regression between them. These results showed that $q=0.53$ which is an identical result as that of \citet{Rauw}, which is unsurprising as the majority of the data were also presented by \citet{Rauw}. 

With the orbital elements and the assumed semi-amplitude of the O star from the mass ratio, we were able to then infer masses for the O and WR stars to be 26.3 and 13.9 $M_\odot$ respectively, with errors shown in Table \ref{OrbitalElements}. The resulting orbital parallax is 0.469 mas, corresponding to a distance of 2.132$\pm$0.054 kpc. We note that the distance according to the Bayesian inference of the Gaia EDR3 measurements is 2.134$^{+0.115}_{-0.093}$ kpc according to \citet{2021AJ....161..147B}, providing confidence in our results.

\section{Discussion}

With our orbit, we can begin to explore the system and how it relates to other WR and O stars. We will begin the comparisons by examining how the system compares to other WN+O binaries as well as the masses from theoretical expectations. The Wolf-Rayet star in WR 138 has the same spectral type as the only other WN star with a visual orbit, WR 133 \citep{Richardson2021}, namely WN5o. In the WR 133 system, \citet{Richardson2021} found the mass of the WN star was 9.3$\pm$1.6 $M_\odot$, which is considerably smaller than the WR star in WR 138 measuring 13.9$\pm$1.5 $M_\odot$.

A better comparison to WN stars with measured masses could be made with short-period binaries with either geometric or wind eclipses. The largest sample of somewhat similar spectral types of WN stars is in the photometric analysis of \citet{1996AJ....112.2227L}, who modeled the wind eclipses in a sample of eleven short-period systems. The wind eclipses are caused by electon scattering as the WR star passes in front of the O star, with the ionized wind of the WR star having the free electrons to scatter the light of the O star. The resulting `v'-shaped eclipses are then dependent on orbital parameters measured from spectroscopy (e.g., $P$, $T_0$, $a \sin i$), along with the mass-loss rate of the WR star (free electrons), and the orbital inclination. In the cases of the WN4 and WN6 stars measured by \citet{1996AJ....112.2227L}, they had masses in the range of 15--19 $M_\odot$, which is quite similar to our measurement of 13.9 $M_\odot$.

We can also use the spectroscopic model for the system reported by \citet{Richardson2016}. With the modeled parameters, we can then use mass-luminosity relations such as those of \citet{2011A&A...535A..56G}, which would place the WR star at 12.8 $\pm$0.5 $M_\odot$, close to that of our measurement. We do caution that this value is dependent on the luminosity of the star, which \citet{Richardson2016} placed at a distance of 1.38 kpc, but our visual orbit and the \textit{Gaia} measurements have both placed at a distance of 2.1 kpc. It is beyond the scope of this paper to recalculate the spectroscopic models of the system, but the general agreement of the WR mass with similar WR stars and the predictions of the mass-luminosity relations is promising. 

For the O star, there are two plausible routes for comparing the measured mass of 26.3$\pm$1.7 $M_\odot$ to other O stars. First, we will use the spectroscopic models of the binary from \citet{Richardson2016} again. In this case, the constraints come from the measured values of $\log g$ (cgs), which was 4.0$\pm$0.3 dex. Unfortunately, the large error on this parameter means that the O star mass from the spectroscopic models was 29$\pm$19 $M_\odot$, which obviously agrees with our value given the error in the spectroscopic models. 

Another way to consider the mass of the secondary is to use its spectral type and the models for the O star masses, namely those of \citet{2005A&A...436.1049M}. \citet{Richardson2016} report the spectral type of the companion to be an O9V star. An O9V star should have a mass around 18 $M_\odot$, which is a bit lower than our value of 26 $M_\odot$. However, an O9III star would have a mass close to 23 $M_\odot$, very similar to our measured value. We note that the spectral luminosity classification of O9 stars is largely done with weak lines in the blue such as \ion{Si}{4} $\lambda \lambda$4089,4116 and \ion{N}{5} $\lambda$4379. These lines are weak in all luminosity classes, and would likely be very hard to detect with the combined spectrum of a WN star and a projected rotational velocity of the O star of 350$\pm$30 km s$^{-1}$ \citep{Richardson2016}. 

\citet{Palate} examined both the radial velocity and X-ray variability of the WR\,138 system. The X-ray observations of WR\,138 are sparse and taken with multiple satellites. The models of the six epochs of observations show some variation, but the largest variation was seen with \textit{ROSAT} which had a very small energy range for which it was sensitive. It is likely that the system should show a variation dependent on the separation $D$ in the system, resulting in either an adiabatic cooling ($D^{-1}$) or a radiative cooling dependency ($D^{-2}$) as described by \citet{1996ApJ...469..729C} and \citet{2009ApJ...703...89G}. The observations presented by \citet{Palate} are not dense enough in phase coverage nor all of high enough quality for an appropriate fit of the cooling of the gas. We suggest that a dedicated X-ray variability campaign across an entire orbit of WR\,138 should be a high-priority in order to best constrain the variability of the system and place constraints on how the wind collisions cool in orbits with well-established orbits.

Other observations of WR\,138 could allow for better constraints on the colliding wind geometry and a better understanding of the way in which polarization is impacted by the geometry of the colliding winds. \citet{2020AJ....159..214F} examined many WR+O binaries with spectropolarimetry, finding a fairly small polarization for WR\,138. The SMEX satellite Polstar is currently being proposed to NASA as a small mission to explore the wavelength-dependent polarization of stars in the ultraviolet. Compared to the mission expectations, WR\,138 is fainter than most of the main targets. However, with selected epochs to observe the system and long ($\sim$1 day) exposures, strong constraints on the polarization changes could provide insights into the wind collisions given the known orbital elements presented here \citep{2022Ap&SS.367..118S} as well as the rapid rotation of the O star companion \citep{2022Ap&SS.367..124J}.

\begin{figure}
    \centering
    \includegraphics[width=0.9\columnwidth]{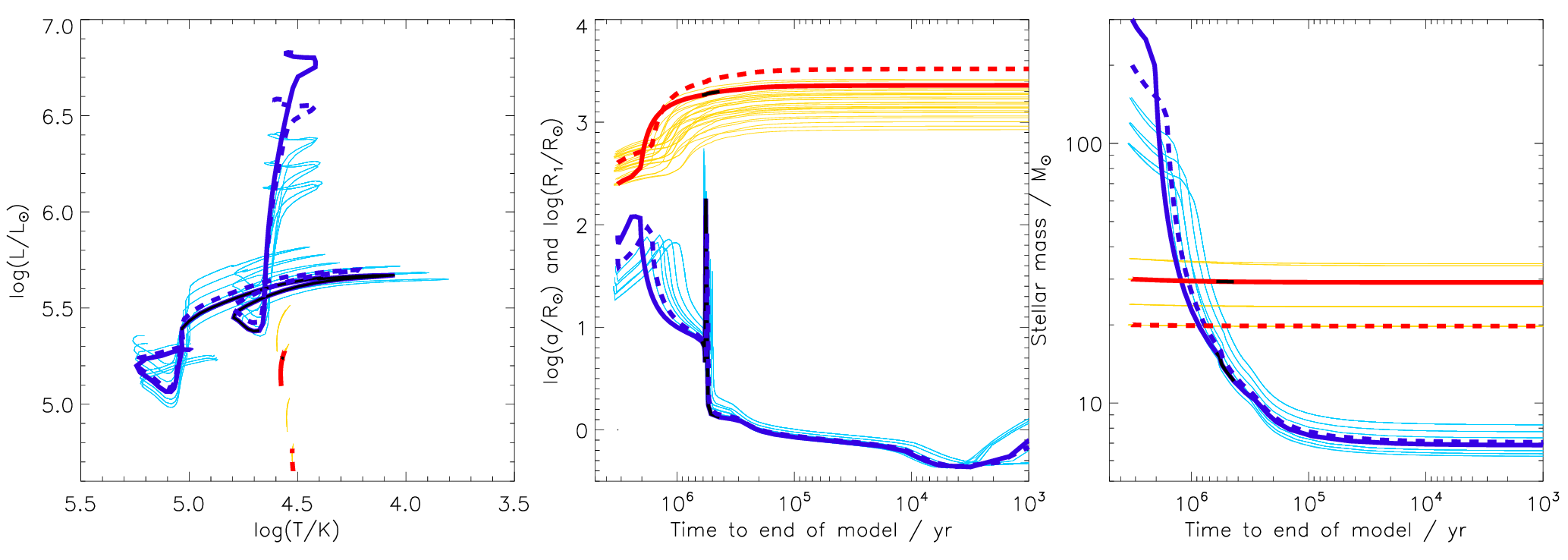}
    \includegraphics[width=0.9\columnwidth]{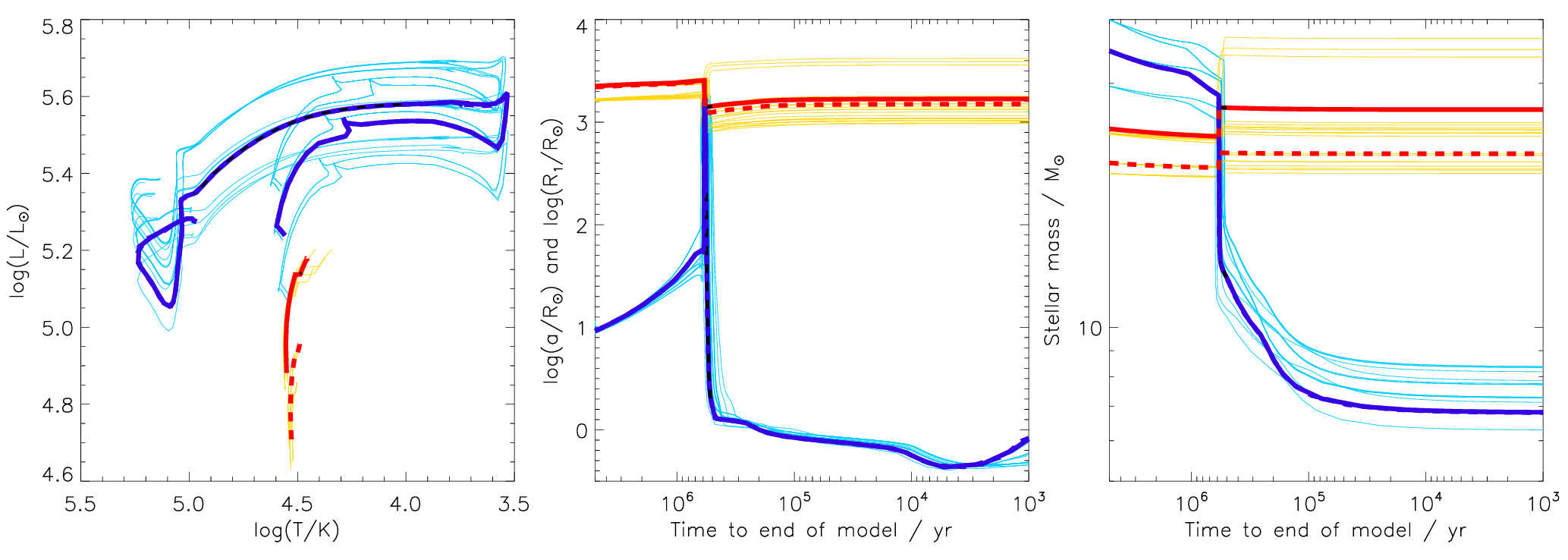}
    \caption{Different aspects of evolution of the WR\,138 system are shown in these three panels. In each of the figures the blue and red bold lines represent the model with the best matching initial parameters with thinner lined models that are match the observed masses and period within 3$\sigma$ uncertainties. The mean model is shown as a thick dashed line while the mode model is shown as a thick solid line, note in the lower panels the mean and mode models are almost identical. The lower panels for the lower initial mass evolutionary pathway while the upper panels are for the higher initial mass evolutionary pathway for WR138. In the left panels, we show the Hertzsprung-Russell diagram for the past and future evolution of the WR star. In the central panels, we show the primary radius in light/dark blue and the orbital separation in yellow/red. In the right panels we show the mass of the primary in light/dark blue and the mass of the secondary in yellow/red.}
    \label{fig:evolution}
\end{figure}

{ We compare the observational constraints for WR\,138, including the current masses, the $UBVJHK$ magnitudes and current circularized separation of the binary to BPASS v2.2 binary stellar evolution models \citep{2017PASA...34...58E, 2018MNRAS.479...75S} at over the range of metallicities allowed in BPASS.
The comparison to BPASS models reveals two main evolutionary pathways sets of models that agree with the current observed parameters which we show in Fig.~\ref{fig:evolution}. The first pathway has initial masses for the binary system of 178$\pm$71~M$_{\odot}$ and 28.2$\pm$2.8~M$_{\odot}$ with initial periods of $\log(P/{\rm days}) = 1.76\pm0.26$ with a super-Solar metallicity mass fraction of $Z=0.038\pm0.005$. These systems do not interact and the primary star loses most of its mass by strong stellar winds on the main sequence, before any binary interaction can occur.  This mass loss is what drives the system to such long observed periods. The current age predicted today is 2.88~Myrs.

The second pathway also shown in Fig.~\ref{fig:evolution} has initial masses of the binary systems with of 37.4$\pm$7.5~M$_{\odot}$ and 24.0$\pm$1.4~M$_{\odot}$, initial periods of $\log(P/{\rm days})$ = 3.17$\pm$0.16 and a slightly lower metallicity of $Z=0.026\pm0.009$. These systems interact when the primary reaches its red supergiant phase. The interactions begin with some slight mass transfer before a common envelope phase reducing the orbit slightly to values observed today. The current age predicted today is 5.0~Myrs. Both sets of models will have the current O star accreting some material and angular momentum, aiding the star to becoming the rapid rotator we observe today, even though rotation is not fully accounted for in BPASS. 

When considering the full set of models the best fitting initial parameters have an initial primary mass of 35~M$_{\odot}$, a secondary mass of 24.5~M$_{\odot}$ and an initial period of $\log(P/{\rm days}) = 3.2$. While the higher mass model set are possible, we consider that the lower mass set is more representative of the binary system's prior evolution, especially with the steep initial mass function for star formation. As discussed in the BPASS modeling of WR\,140 \citep{2021MNRAS.504.5221T}, the caveats around these fits is that BPASS is currently unable to model eccentric orbits. However, as discussed by \citet{2002MNRAS.329..897H}, orbits with the same semi-latus rectum evolve through similar pathways. Thus we have constrained our models to have the same circular orbital radius as the semi-latus rectum of the observed binary. Given the relatively low value of eccentricity, we expect this to be a good approximation in this case.

WR\,138 is a member of a growing class of massive, eccentric binaries with observational and theoretical evidence of at least some binary interactions being necessary to form the system as observed today. Such systems have included the LBV candidates HD 326823 and MWC 314 \citep{2011AJ....142..201R, 2016MNRAS.455..244R} and even the prototype LBV binary $\eta$ Car \citep{2021MNRAS.503.4276H}. With $\eta$ Car, the models indicate a merger that formed the modern day primary star after complex interactions with a third component. However, in the case of the short period systems HD\,326823 and MWC\,314, the eccentricity observed in these interacting binaries is hyopthesized to be driven through a transfer of angular momentum during a periastron passage that works to increase eccentricity with time. The process of increasing the eccentricity with time was modeled in a series of papers by \citet{2007ApJ...660.1624S, 2007ApJ...667.1170S, 2009ApJ...702.1387S, 2010ApJ...724..546S}. Some evidence presented in the evolutionary analysis of WR\,140 presented by \citet{2021MNRAS.504.5221T} also indicated that the high eccentricity (0.9) was driven by the angular momentum transfer at periastron increasing the eccentricity with time. If similar results are seen now with WR\,138, it is an opportune time for theorists to model how these interactions can occur to build a growing number of eccentric binaries with both short and long periods. }

\section{Conclusions}
We have presented the second visual orbit for a WN-type star in a binary system derived using a combination of long-baseline infrared interferometry and radial velocities from optical spectra. The resulting masses are in agreement with the masses expected from spectral modeling previously done for this system and the orbital parallax derived is in agreement with the \textit{Gaia} parallax. The observations reported here show that the speculation of a third body in the system by \citet{Rauw} is not plausible { although we adjusted different data sets' $\gamma$-velocity to fit our orbit. We suspect that this adjustment only adjusts the various measuring techniques from different authors more than an intrinsic change in the $\gamma$-velocity with time, and furthermore we see no evidence of a third body in our interferometry}. Furthermore, the system may have undergone some past interactions through a common envelope phase or mass transfer when the current WN star was in a red supergiant phase. Finding more systems like WR\,138 that can be measured with both spectroscopy and interferometry will place strong constraints on the formation mechanisms for these stars and binaries in the future.

\section*{acknowledgments}

We thank Peredur Williams for comments that improved this manuscript. This work is based upon observations obtained with the Georgia State University Center for High Angular Resolution Astronomy Array at Mount Wilson Observatory. The CHARA Array is supported by the National Science Foundation under Grant No. AST-1636624 and AST-2034336.  Institutional support has been provided from the GSU College of Arts and Sciences and the GSU Office of the Vice President for Research and Economic Development. Time at the CHARA Array was granted through the NOIRLab community access program (NOIRLab PropIDs: 2017B-0088, 2021B-0159, and 2023A-452855; PI: N. Richardson). This research has made use of the Jean-Marie Mariotti Center Aspro and SearchCal services.

Some of the data presented herein were obtained at the W. M. Keck Observatory, which is operated as a scientific partnership among the California Institute of Technology, the University of California and the National Aeronautics and Space Administration. The Observatory was made possible by the generous financial support of the W. M. Keck Foundation.
The authors wish to recognize and acknowledge the very significant cultural role and reverence that the summit of Maunakea has always had within the indigenous Hawaiian community.  We are most fortunate to have the opportunity to conduct observations from this mountain.

AMH is grateful for support through Embry-Riddle Aeronautical University's Undergraduate Research Institute and the NASA Arizona Space Grant program. NDR is grateful for support from the Cottrell Scholar Award \#CS-CSA-2023-143 sponsored by the Research Corporation for Science Advancement. SK acknowledges funding for MIRC-X received funding from the European Research Council (ERC) under the European Union's Horizon 2020 research and innovation programme (Starting Grant No. 639889 and Consolidated Grant No. 101003096). JDM acknowledges funding for the development of MIRC-X (NASA-XRP NNX16AD43G, NSF-AST 1909165) and MYSTIC (NSF-ATI 1506540, NSF-AST 1909165). JM acknowledges funding from a Royal Society - Science
Foundation Ireland University Research Fellowship.


\vspace{5mm}
\facilities{CHARA}

\software{astropy \citep{2013A&A...558A..33A,2018AJ....156..123A}, MIRC-X reduction software \citep{le_bouquin_2024_12735292}}

\appendix
The appendix includes Figure Set 4 showing the interferometric data and the binary fits for each epoch of MIRC-X and MYSTIC. Each figure in the set shows the ($u,v$) coverage, the $\chi^2$ map from the binary grid search, the visibilities, and the closure phases. The $\chi^2$ maps are centered at the predicted location based on the updated orbit fit. The nights with reliable detections show a clear minimum in the $\chi^2$ indicated by the colored circles. { The large red, orange, yellow, green, blue, purple, and black symbols show solutions within a $\Delta chi^2$ interval of 1, 4, 9, 16, 25, 36, and 49 from the minimum $\chi^2$. The small black circles show solutions where the difference in the $\chi^2$ is greater than 49. Similar plots are shown in the appendix of Richardson et al.~(2024) and show that the high-quality interferometric data from MIRC-X and MYSTIC frequently will only show the best fit, and hence only red points}.

{\bf NOTE for arXiv readers: These figures make the pdf too large for arXiv and we only include one. A full copy can be requested from the corresponding author until the paper is published in ApJ.}



\begin{figure}
\figurenum{4.1}
  \begin{center}
	\scalebox{0.42}{\includegraphics{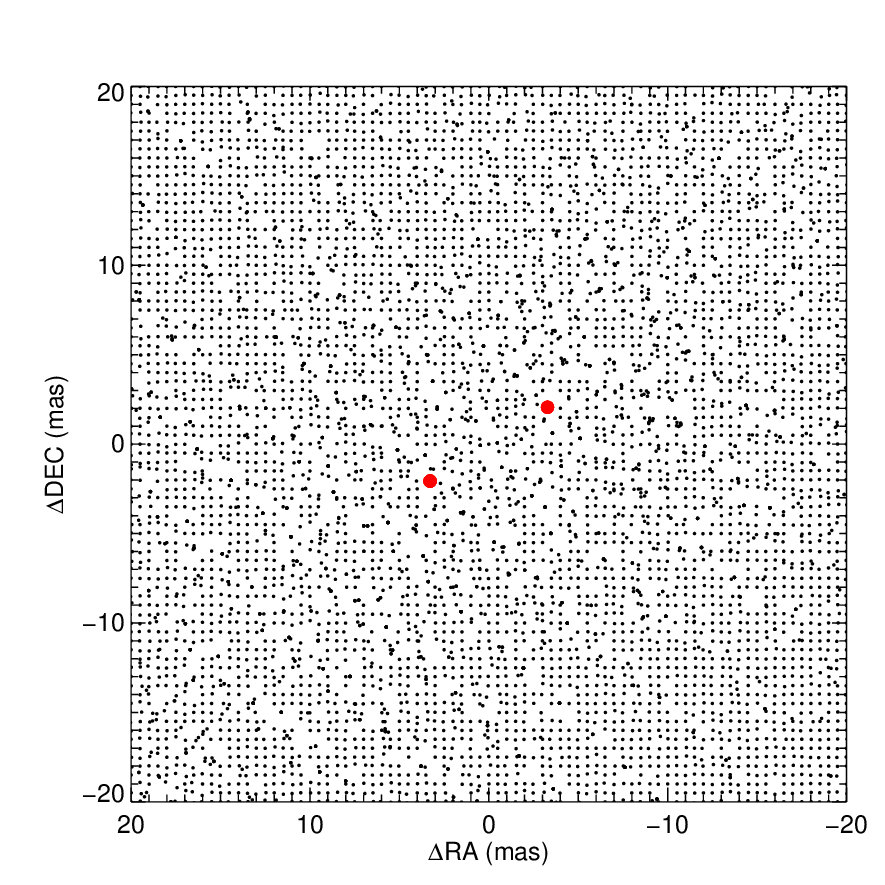}}
	\scalebox{0.54}{\includegraphics{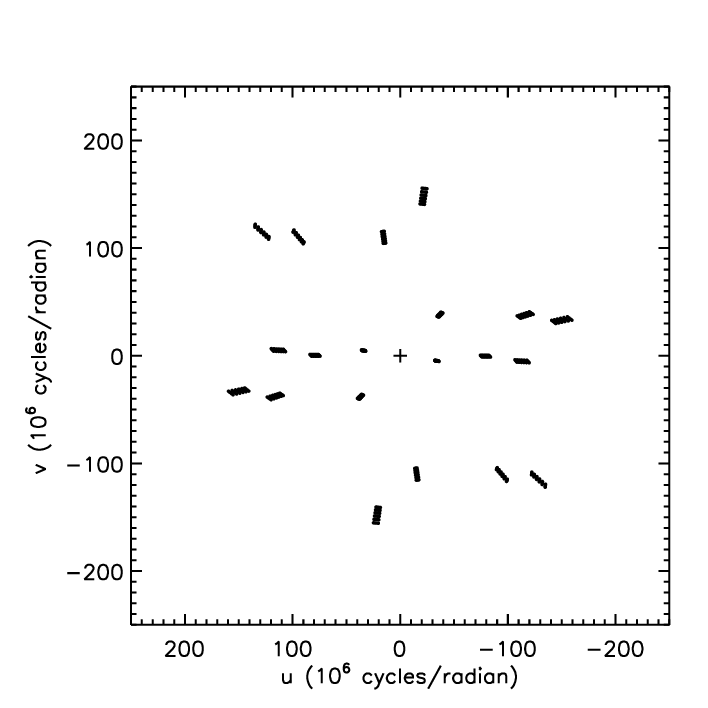}} \\
	\scalebox{0.33}{\includegraphics{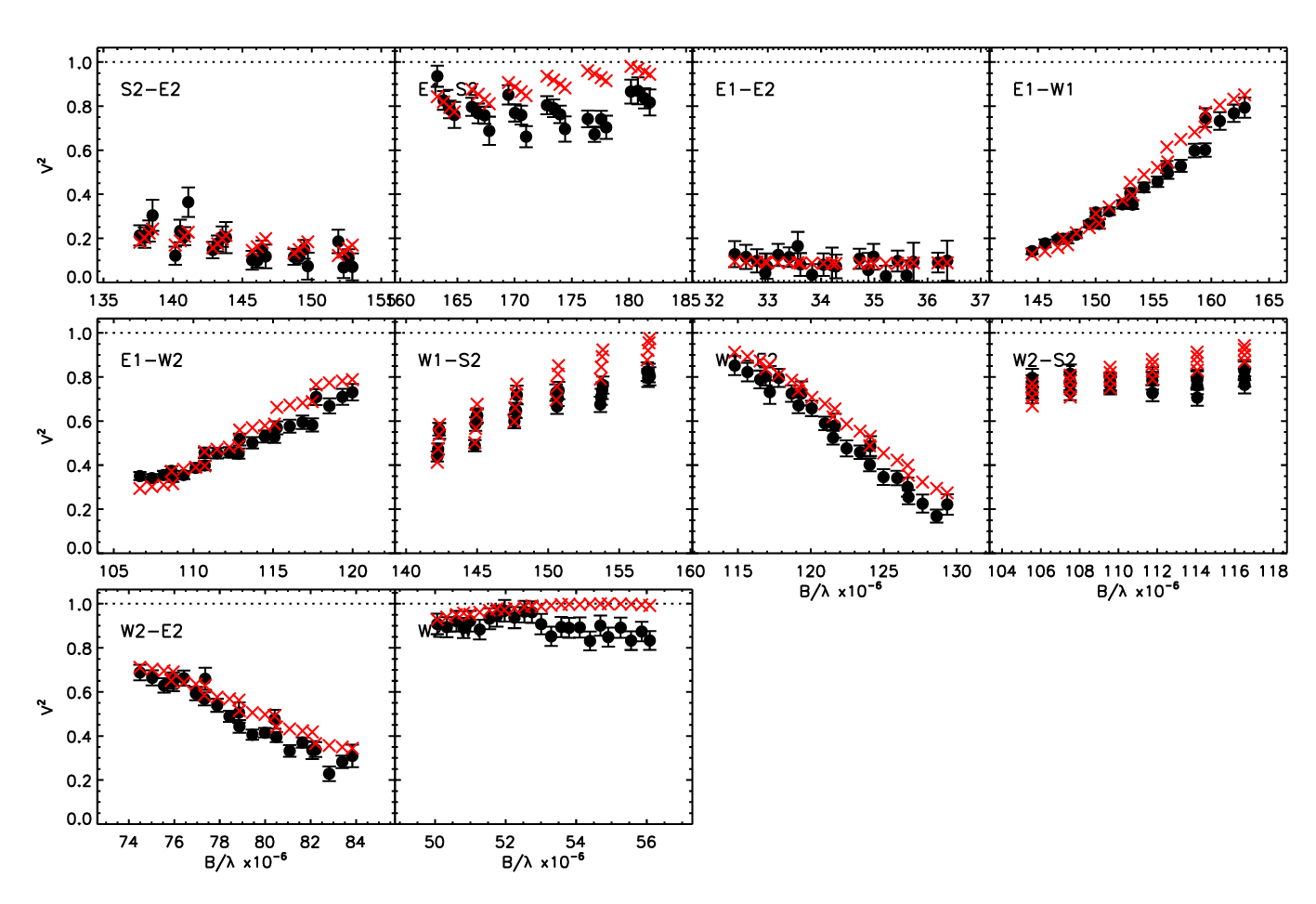}}
	\scalebox{0.33}{\includegraphics{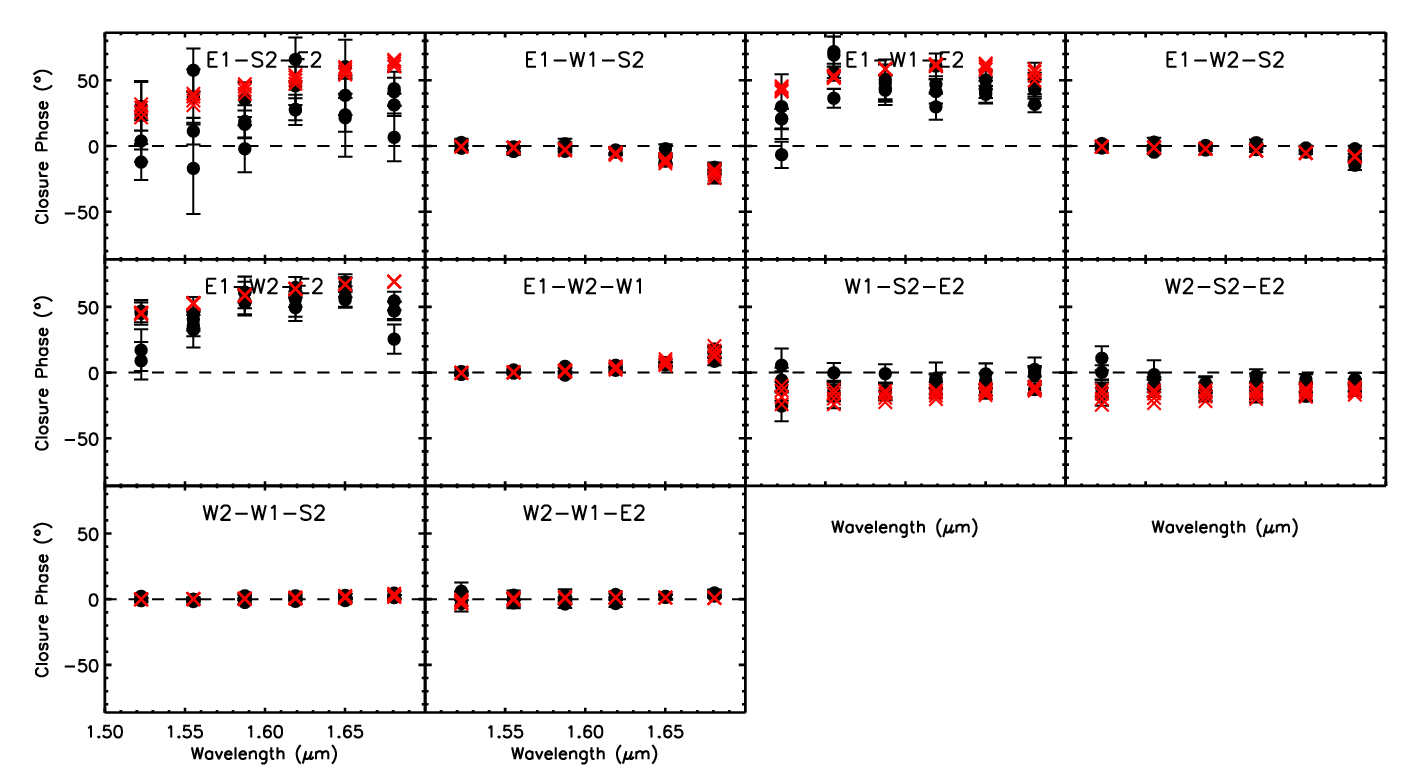}}
        \caption{Top row: $\chi^2$ map (left) from binary fit for WR 138 and $uv$ coverage (right) for MIRC-X data obtained on UT 2019Jul01 (set1).  Bottom row: Visibilities (left) and closure phases (right). Black circles - measured values.  Red crosses - binary fit.}
  \end{center}
\end{figure}

\bibliography{sample631}{}
\bibliographystyle{aasjournal}

\end{document}